\documentclass[preprint]{elsarticle}

\newcommand{\mathbb}[1]{\mathbf{#1}}
\newcommand{\idx}[1]{_{\mathrm{#1}}}

\newcommand{\vs}{\textit{vs}}
\newcommand{\via}{\textit{via}}

\begin{document}

\title{Stationary states and spatial patterning in an $SIS$ epidemiology model
with implicit mobility}

\author[ji]{J.M.~Ilnytskyi\corref{cor}}\ead{iln@icmp.lviv.ua}
\author[yk]{Y.~Kozitsky}
\author[hi]{H.I.~Ilnytskyi}
\author[oh]{O.~Haiduchok}
\cortext[cor]{Corresponding author}

\address[ji]{Institute for Condensed Matter Physics of Nat. Acad. Sci. of Ukraine, Lviv, Ukraine}
\address[yk]{Maria Curie-Skłodowska University, Lublin, Poland}
\address[hi]{Danylo Halytsky Lviv National Medical University, Lviv, Ukraine}
\address[oh]{Lviv Polytechnic National University, Lviv, Ukraine}
 
\date{\today}

\begin{abstract}
By means of the asynchronous cellular automata algorithm we study stationary states and
spatial patterning in an $SIS$ model, in which the individuals' are attached to the
vertices of a graph and their mobility is mimicked by varying the neighbourhood size $q$.
The versions with fixed $q$ and those taken at random at each step and for each individual are studied. Numerical data on the local behaviour of the model are mapped onto the solution of its zero dimensional version, corresponding to the limit $q\to +\infty$ and equivalent to the logistic growth model. 
This allows for deducing an explicit form of the dependence of the fraction of infected individuals on
the curing rate $\gamma$. A detailed analysis of the appearance of spatial patterns of
infected individuals in the stationary state is performed.
 
\end{abstract}


\maketitle

\section{\label{I}Introduction}

As the spread of a severe disease affects large number of individuals its deep understanding calls for the use of mathematical models of complex dynamical systems. 
In the models of the Kermack-McKendrick type \cite{Kermack1927}, the population is split into characteristic
groups (compartments)  such as: susceptible to infection; infected; removed (recovered). The corresponding
fractions are traditionally denoted as $S$, $I$ and $R$, respectively. Their evolution is obtained from
a system of differential equations which take into account the basic mechanisms of curing and infecting.
Depending on the compartments involved the models are traditionally abbreviated as $SIR$ and $SIS$ (for the
case of non-immune disease where $R$ group is absent). Further refinements are made by introducing such
compartments as latently infected $E$, or by splitting existing compartments into sub-compartments based on
such  differentiations as age, sex, immunity level \cite{Brauer2005,Sun2010,Hu2012}.
Usually, such models do not take into account the spatial localization of the individuals involved in the process,
which is adequate only if the described populations are well-mixed and hence behave globally in space. 
That is why, such models are often referred to as zero-dimensional.

A more advanced description should, however, be based on the local properties of the system and on the deduction of the global properties from the local ones. This includes describing the appearance of spatial patterns
(clustering). Along with analytic methods spatially-structured models are also studied by means of the
cellular automata (CA). Here one considers a system of finite collections of individuals. Each individual
is attached to a vertex of an underlying graph.  The evolution of the system is then governed by a
specific update algorithm which changes the state of a particular individual according to the states of its
neighbours in the graph. This type of evolution is discrete in space and time and may involve explicit or
implicit account for the mobility of its constituents
\cite{Boccara1992,Boccara1993,Boccara1994,Ahmed1998,Fuentes1999,Kuperman1999,Sirakoulis2000,Fuks2001,Fu_Milne2003,Situngkir2004,Hiebeler2005,Beauchemin2005,White2007}.
Boccara and Cheung \cite{Boccara1992} considered $SIR$ CA model with
two subrules: for explicit individuals mobility and for the infecting/curing update. They studied the influence of
the degree of mixing due to the mobility on the epidemics spread and found, in particular, that for an infinity
degree of mixing the time evolution turns into that predicted by the zero dimensional $SIR$ model.
Then  the same authors extended their study to the case of CA $SIS$ model  \cite{Boccara1993}. A kind of phase
transition was found between the endemic and disease-free states, where the role of the order parameter is
played by the stationary value of $I$, dependent on the model parameters.
It was found that the mobility essentially determines the behaviour of the model. In particular,
it reduces to that of the zero-dimensional $SIS$ model in the limit of large number of tentative moves,
see also \cite{Boccara1994}. Another way of accounting mobility is as follows. 
The individuals are set static and their mobility is mimicked by varying the neighbourhood size $q$. 
This approach is used in the current study in two versions: (a) the neighbourhood size is fixed;
(b) it is random at each time step and for each individual. The latter allows for mimicking also the
spatial heterogeneity of infectivity as studied in \cite{Fuks2001,German2011}. 
In the study of dynamical systems, one usually finds the stationary state and then describes
how the system approaches this state. Periodic temporal modulations of parameters corresponding to some
epidemic features are considered
in Ref.~\cite{Kuperman1999} for CA $SIS$ model, it was found that the inclusion of the vital dynamics
introduces no relevant changes in its behaviour.

In the present work, we consider CA $SIS$ model on geometric graph obtained from 
$\mathbb{Z}^2$ by assigning neighbours to a given $k \in \mathbb{Z}^2$ according to the rule:
$k' \sim k$ whenever $|k'-k|\leq R_q$ for a given $R_q>0$, where $R_q$ is such that the neighbourhood
size is equal to $q$. The latter paremeter can be either fixed or set random. The aim of the study
is
\begin{itemize}
\item for a fixed $q$, studying of the ratio of infected individuals in the stationary state and its
dependence on the model parameters;
\item the same for globally bounded random $q$ analysing the role of randomness;  
\item spatial patterning and dynamics of approaching the stationary state.
\end{itemize}

In Sec.~\ref{II}, we present the solution for the zero dimensional $SIS$ model in the form
used in the remainder of the paper.
In Sec.~\ref{III}, we consider our main model with the fixed neighbourhood
and relate it to that of the zero dimensional version with rescaled infectivity. In Sec.~\ref{IV},
we study version of the model in which the neighbourhood of each vertex is random of globally bounded size.
In Sec.~\ref{V}, we discuss the spatial patterning and the time evolution towards the stationary state.
Conclusions are made in Sec.~\ref{VI}. 

\section{\label{II}Zero dimensional $SIS$ model}

\begin{figure}
\begin{center}
\includegraphics[clip,width=4cm]{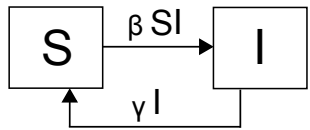}
\caption{\label{SIS_model}Flow chart for the zero dimensional $SIS$ model.}
\end{center}
\end{figure}
The flow chart for the zero dimensional $SIS$ model \cite{Kermack1927} which we consider is shown in
Fig.~\ref{SIS_model}. Its evolution is governed by the differential equations
\begin{equation}\label{SIS}
\left\{
\begin{array}{l}
\dot{S}=\gamma I - \beta I S\\
\dot{I}=-\gamma I + \beta I S,
\end{array}
\right.
\end{equation}
where dot stands for the time derivative and $\beta$ and $\gamma$ are infecting and curing rates, respectively.
By the condition $S=1-I$ the system turns into
\begin{equation}\label{kin_SIS}
\dot{I}=(\beta-\gamma) I - \beta I^2=(\beta-\gamma)I\left[1-\frac{I}{1-\gamma/\beta}\right],
\end{equation}
which is the logistic growth equation \cite{Wolfram2002,Perthame2007}. In the sequel, as a
reference model we use the reduced version of (\ref{kin_SIS}) corresponding to the choice
\begin{equation}\label{SIS_red_relation}
\beta=1-\gamma > \gamma,
\end{equation}
which amounts to considering $\gamma$ restricted to $\gamma<0.5=:\gamma_c$. In this case,
\begin{equation}\label{SIS_red_infty}
I(\infty)=\frac{1-\gamma/\gamma_c}{1-\gamma}
\end{equation}
and the explicit solution is given by
\begin{equation}\label{SIS_red_solut}
I(t)=\frac{I(0)I(\infty)}{I(0)+[I(\infty)-I(0)]e^{-t/\tau}},
\end{equation}
where $\tau=1/(1-\gamma/\gamma_c)$ is a characteristic time for the system to reach the stationary state.

\section{\label{III}Cellular automaton $SIS$ model with fixed neighbourhood size}

We consider the CA $SIS$ model on the following geometric graph. The vertex set is $\mathbb{Z}^2$,
and, for a fixed $R_q>0$, the neighbourhood of a given $k \in \mathbb{Z}^2$ is defined according to the rule:
$k' \sim k$ whenever $|k'-k|\leq R_q$. The choice of $R_q$ is made to obtain the neighbourhood size
equal $q$. To each $k$ there is attached a individual, the state of which, $s_k$, is either $0$ (susceptible) or
$1$ (infected). The evolution of the system of $N$ such individuals is run according to the algorithm
\begin{itemize}
 \item select individual $k$ at random;
 \item if $s_k=0$ do nothing;
 \item if $s_k=1$ then set $s_k=0$ with probability $\gamma$ and set
  $s_{k'}=1$ for one of its randomly chosen neighbours with probability $1-\gamma$;
 \item perform $N$ attempts described as above to complete a single time-step.
\end{itemize}
It is asynchronous as far as the flip of $s_k$ is immediate and it affects the updates of other individuals
before the time-step is complete. On average, the probability $p(1 \rightarrow 0)$ for infected individual to be
cured and for the susceptible individual to be infected, $p(0 \rightarrow 1)$, are equal to:
\begin{eqnarray}\label{probs}
p(1 \rightarrow 0)&=&\gamma,\\
p(0 \rightarrow 1)&=&(1-\gamma)\,i_k/q,\nonumber
\end{eqnarray}
where $i_k$ is the number of infected individuals in the neighbourhood of $k$th individual, therefore
the infecting is now local.

In the relation to this, one can refer to the following specific cases: $q=4$ corresponds
to CA $SIS$ with the von Neumann neighbourhood, referred also to as the contact process on a square lattice
\cite{Harris1974}; $q=8$ realises the so-called Moore neighbourhood; at $q=N$ the
infecting range spans over all the system representing the zero dimensional $SIS$ model (\ref{SIS})
considered in the previous section.

\begin{figure}
\begin{center}
\includegraphics[clip,angle=270,width=8cm]{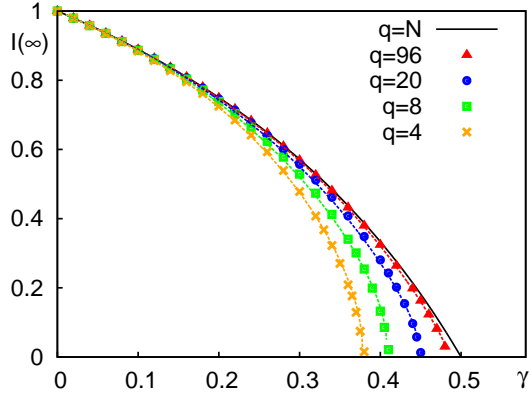}
\caption{\label{qfix_I}Fraction $I(\infty)$ of infected individuals in the stationary state
\vs\ curing rate $\gamma$ for the CA $SIS$ model with fixed neighbourhood size $q$.
Symbols: simulation data; dashed curves: Eq.~(\ref{SIS_red2_solut}) using power law fit
(\ref{theta_form}); black solid curve: Eq.~(\ref{SIS_red_infty}) for the zero dimensional
$SIS$ model. Values for critical curing rate $\gamma_c$ are: $0.378$ ($q=4$), $0.408$ ($q=8$),
$0.447$ ($q=20$), $0.482$ ($q=96$), and $0.5$ ($q=N$).}
\end{center}
\end{figure}
All simulations in the current study are performed on the square of $256 \times 256$ individuals,
total number of individuals is $N=65536$. The periodic boundary conditions are applied in both directions.
Typical times for the system to reach the stationary state vary considerably depending on the value of
$I(0)$ and on the proximity to the critical value $\gamma_c$, where the
relaxation time is the longest. The value of $\gamma_c$ depends on $q$ and is
defined by the condition that the fraction of infected individuals vanish.
Near $\gamma_c$, $I(\infty)\sim 0.1$, hence the choice of $I(0)=0.1$ reduces drastically the time
needed by the system to reach the stationary state when $\gamma\approx\gamma_c$. This choice for
$I(0)$ is used in our study and the simulations of $2000$ time-steps are performed at each $\gamma$
ranging from $\gamma=0$ to $\gamma\approx\gamma_c$. To avoid locking the system in the state with
$I(t)=0$, we allow for at least one infected individual in the system.
The dependence of $I(\infty)$ \vs\ curing rate $\gamma$ at various $q$ is shown in Fig.~\ref{qfix_I}.
At $q=N$, it follows the solution (\ref{SIS_red_solut}). For $q<N$, the general shape is the same but
with reduced critical curing rate $\gamma_c$.
At $q=4$, we obtain $\gamma_c\approx 0.368$, which agree well with the value $\lambda_c=1.64872(3)$
found by Sabag {\it et al.} \cite{Sabag2002}, where $\lambda_c=(1-\gamma_c)/\gamma_c$.

The CA $SIS$ model is characterised by the local infectivity $i_k/q$ (\ref{probs}), in contrary
to the zero dimensional $SIS$ model, in which case the infectivity is global and equal to $I$.
Using the simulation data obtained at $q<N$, the zero dimensional $SIS$ model can be extended to the
case of $q<N$ by the following substitution for the infecting rate: $\beta\rightarrow\beta'$, where
%
\begin{displaymath}
\beta'=\beta\frac{\langle i_k/q_k\rangle}{I}
\end{displaymath}
%
If the averaging is performed in the stationary state, then the dependence of the scaling factor for
$\beta'$ on $I$ is eliminated. Taking also into account (\ref{SIS_red_relation}), one obtains
\begin{equation}\label{beta_theta}
\beta'=\beta\theta(\gamma),\;\;\theta(\gamma)=\frac{\langle i_k/q_k\rangle\idx{stat}}{I}
\end{equation}
where ``stat'' indicates averaging in the stationary state. By solving (\ref{kin_SIS}) with the
substitution of $\beta$ by $\beta'$ according to (\ref{beta_theta}) and taking into account (\ref{SIS_red_relation}),
one obtains $I(\infty)$ for the extended zero dimensional $SIS$ model
\begin{equation}\label{SIS_red2_solut}
I(\infty)=\frac{1-[1+\theta^{-1}(\gamma)]\gamma}{1-\gamma},\hspace{2em} \gamma<\gamma_c.
\end{equation}
It yields the following equation for the critical curing rate $\gamma_c$:
\begin{equation}\label{gammac_equat}
\theta(\gamma_c)=\frac{\gamma_c}{1-\gamma_c}
\end{equation}
which is not equal to $0.5$ now.

\begin{figure}
\begin{center}
\includegraphics[clip,angle=270,width=8cm]{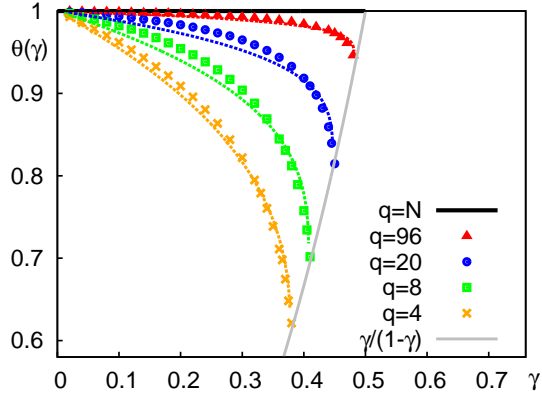}
\caption{\label{qfix_theta}Scaling factor $\theta(\gamma)$ in the stationary state \vs\ curing rate $\gamma$
for the extended zero dimensional $SIS$ model with fixed $q$. Symbols: simulation data,
dashed curves: analytic expression, Eq.~(\ref{theta_form}). Fitted values for the
exponent $\delta$ are: $0.400$ ($q=4$), $0.319$ ($q=8$), $0.258$ ($q=20$) and $0.175$
($q=96$).}
\end{center}
\end{figure}
To made the extended zero dimensional $SIS$ model self-contained one needs certain analytic form for
$\theta(\gamma)$. It can be obtained by fitting the simulation data obtained at various $q$. The
appearance of $\theta(\gamma)$, as shown in Fig.~\ref{qfix_theta}, suggests the power law of the form
$\sim a(1-\gamma/\gamma_c)^\delta + b$. Constants $a$ and $b$ can be found from
the conditions that: (i) at $\gamma=0$ we have the stationary state with $s_k=1$ for all individuals,
hence $\theta(0)=1$; (ii) at $\gamma=\gamma_c$ the value of $\theta(\gamma_c)$ is given by
Eq.~(\ref{gammac_equat}). Therefore, the only fitting parameter left is the exponent $\delta$.
The fitting expression is
\begin{equation}\label{theta_form}
\theta(\gamma)=\frac{1-2\gamma_c}{1-\gamma_c}
\left(1-\frac{\gamma}{\gamma_c}\right)^\delta + \frac{\gamma_c}{1-\gamma_c},
\end{equation}
and the results of fitting at selected $q$ are shown in Fig.~\ref{qfix_theta} \via\ dashed lines.

\begin{figure}
\begin{center}
\includegraphics[clip,angle=270,width=8cm]{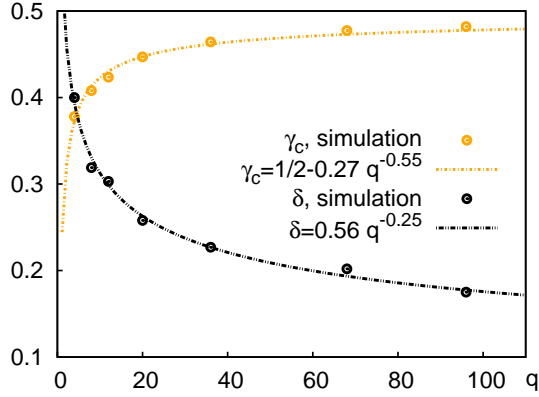}
\caption{\label{qfix_gammac_delta}Dependencies of the critical curing rate
$\gamma_c$ and fitting exponent $\delta$ in Eq.~(\ref{theta_form}) \vs\ $q$ for the
extended zero-dimensional $SIS$ model with fixed $q$.}
\end{center}
\end{figure}
Therefore, the simulations of CA $SIS$ model performed for various neighbourhood size $q$
allow us to extend the zero dimensional $SIS$ model to the case of $q<N$. The solution for
the fraction of infected individuals in the stationary state is given by (\ref{SIS_red2_solut}).
Here the analytic form (\ref{theta_form}) for $\theta(\gamma)$ is found by fitting the
simulation data obtained for CA $SIS$ model. The dependence for the critical curing rate
$\gamma_c$ and the exponent $\delta$ on $q$ can also be modelled from the simulation data,
as shown in Fig.~\ref{qfix_gammac_delta}. Here the following power laws are used:
\begin{equation}\label{gammac_fit}
\gamma_c = \frac12 - A q^{-u},\hspace{2em}
\delta = B q^{-v},
\end{equation}
where $A \approx 0.27$, $u\approx 0.55$, $B \approx 0.56$, and $v \approx 0.25$.

\section{\label{IV}Cellular automaton $SIS$ model with random neighbourhood size}

It has been pointed out that characterisation of each individual by its own intrinsic infectivity
moves the description of the disease spread towards more realistic behaviour \cite{German2011}.
To do so we consider the CA $SIS$ model, where the neighbourhood size $q_k$ is random for each $k$th
individual at each time step. The evolution of the system of $N$ such individuals is run according
to the algorithm
\begin{itemize}
 \item select random individual $k$
 \item if $s_k=0$ do nothing
 \item if $s_k=1$ then set $s_k=0$ with the probability $\gamma$, choose random neighbourood size
  $q_k$ from uniform distribution in $[0;q\idx{max}]$ and set $s_{k'}=1$ for one of its neighbours with the probability $1-\gamma$
 \item perform $N$ attempts described above to complete a single time-step
\end{itemize}
This algorithm is interpreted as the one describing the set of individuals with random implicit mobility.
The random neighbourhood size $q_k$ is globally bounded by $q\idx{max}$.
As far as the critical curing rate $\gamma_c$ decays with the decrease of $q$ (see,
Fig.~\ref{qfix_gammac_delta}), its value for the model with random $q_k<q\idx{max}$ will be lower than
that for the model with fixed neighbourhood size equal to $q\idx{max}$. Both models can be matched
to have the same value of $\gamma_c$, in which case one requires
\begin{displaymath}
\frac{1}{q\idx{max}}\int_{0}^{q\idx{max}}\!\!\!\!\!\!\!\!\gamma_c(q')dq' = \gamma_c(q)
\end{displaymath}
where the l.h.s. represents the average value for $\gamma_c$ for the model with random neighbourhood size
bounded by $q\idx{max}$, whereas the r.h.s. is the same value of $\gamma_c$ obtained for the model
with fixed neighbourhood size at particular $q$. 
Using the analytic fit for $\gamma_c(q)$ (\ref{gammac_fit}) and performing integration we obtain
simple relation between $q\idx{max}$ and matching value of $q$
\begin{equation}\label{qmax_q}
q\idx{max} = q \left[\frac{1}{1-u}\right]^{\frac{1}{u}} \approx 4.27 q.
\end{equation}

\begin{figure}
\begin{center}
\includegraphics[clip,angle=270,width=8cm]{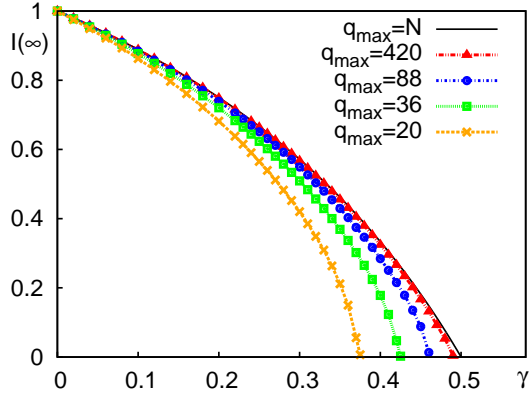}
\caption{\label{qmax_I}The same as in Fig.~\ref{qfix_I} but for the CA $SIS$ model with random
neighbourhood size at various $q\idx{max}$. Values for critical curing rate $\gamma_c$ are: $0.376$ ($q\idx{max}=20$),
$0.426$ ($q\idx{max}=36$), $0.462$ ($q\idx{max}=88$) and $0.491$ ($q\idx{max}=420$).}
\end{center}
\end{figure}

The simulations performed at the values close to those suggested by the relation (\ref{qmax_q}):
$q\idx{max}=20$, $36$, $88$ and $420$ match the respective values for $\gamma_c$ obtained for $q=4$, $8$,
$20$ and $96$ reasonably well, as shown in Fig.~\ref{qmax_I}.

One may conclude that in terms of the critical properties of the stationary state $I(\infty)$, the
CA $SIS$ model with random neighbourhood size can be matched well by its counterpart with appropriately
chosen fixed value of $q$. $q\idx{max}$ and $q$ are related by a simple expression (\ref{qmax_q})
obtained from the fit (\ref{gammac_fit}) for $\gamma_c$ \vs\ $q$. In the next section we will focus
the differences between these two models in terms of the spatial patterns and dynamics of approaching
the stationary state.

\section{\label{V}Spatial patterns and evolution of the system towards the stationary state}

One of principal advantages of using the CA model defined on a graph is the possibility to
analyse spatial patterns of clustered infected individuals, as well as to trace their time evolution
\cite{Fu_Milne2003,Kuperman1999,Athithan2014}. In this study we analyse clustering effects or
infected individuals both in the stationary state and when the system approaches it.
To define clusters we introduce the cluster neighbourhood size, $q_C$, which, in general,
does not coincide with the infecting neighbourhood $q$ or $q_k$. For the case of $\mathbb{Z}^2$
considered here, the value $q_C=4$ is used (the nearest neighbours on the square lattice).
Thereafter, the clusters are enumerated by index $m$ and the cluster label $c_k$ is assigned to
each $k$th individual having the meaning of the host cluster for this individual.
The following algorithm for the clusters recognition is used:
\begin{enumerate}
\item set $\{c_k:=0\}$, $m=0$;
\item pick $k$th individual randomly with $s_k=1$ and $c_k=0$;
\item set $m:=m+1$, $c_k:=m$, declare $k$ as a newcomer;
\item loop over the cluster neighbourhoods $\{l\}$ of all newcomers;
\item if $s_l=1$ then set $c_l:=m$, declare $l$ as a newcomer;
\item repeat steps 4-5 until no newcomers;
\item go to 2.
\end{enumerate}
Alternatively, the Hoshen and Koppelmann \cite{Hoshen1976} algorithm can be used.

At given curing rate $\gamma$ and time instance $t$ we analyse the number of clusters
$N_C(\gamma,t)$ and a set of their sizes $\{S_m(\gamma,t)\}$.
Cluster size is given by the number of infected individuals it contains.
Reduced number of clusters, as well as average and maximum cluster sizes at given $t$
are given as:
\begin{eqnarray}
n_C(\gamma,t) &=& \frac1N N_C(\gamma,t),\nonumber\\
a_C(\gamma,t) &=& \frac1N \langle S_m(\gamma,t) \rangle_m,\nonumber\\
m_C(\gamma,t) &=& \frac1N \max\{S_m(\gamma,t)\}\nonumber.
\end{eqnarray}
Their time averages performed in the stationary state are denoted as $n_C(\gamma)$,
$a_C(\gamma)$ and $m_C(\gamma)$, respectively.

\begin{figure}
\begin{center}
\includegraphics[clip,angle=270,width=8cm]{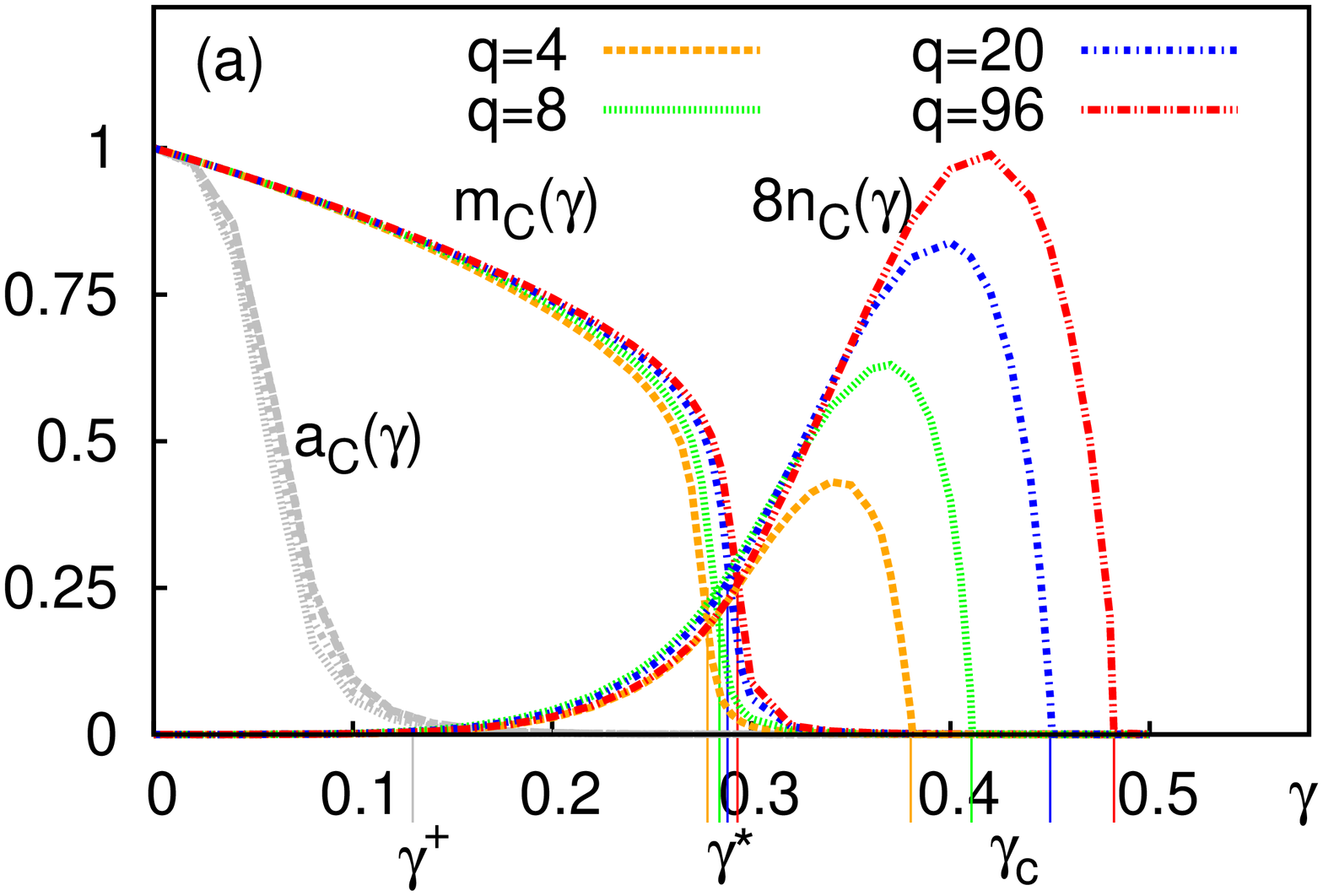}\vspace{-3mm}\\
\includegraphics[clip,angle=270,width=8cm]{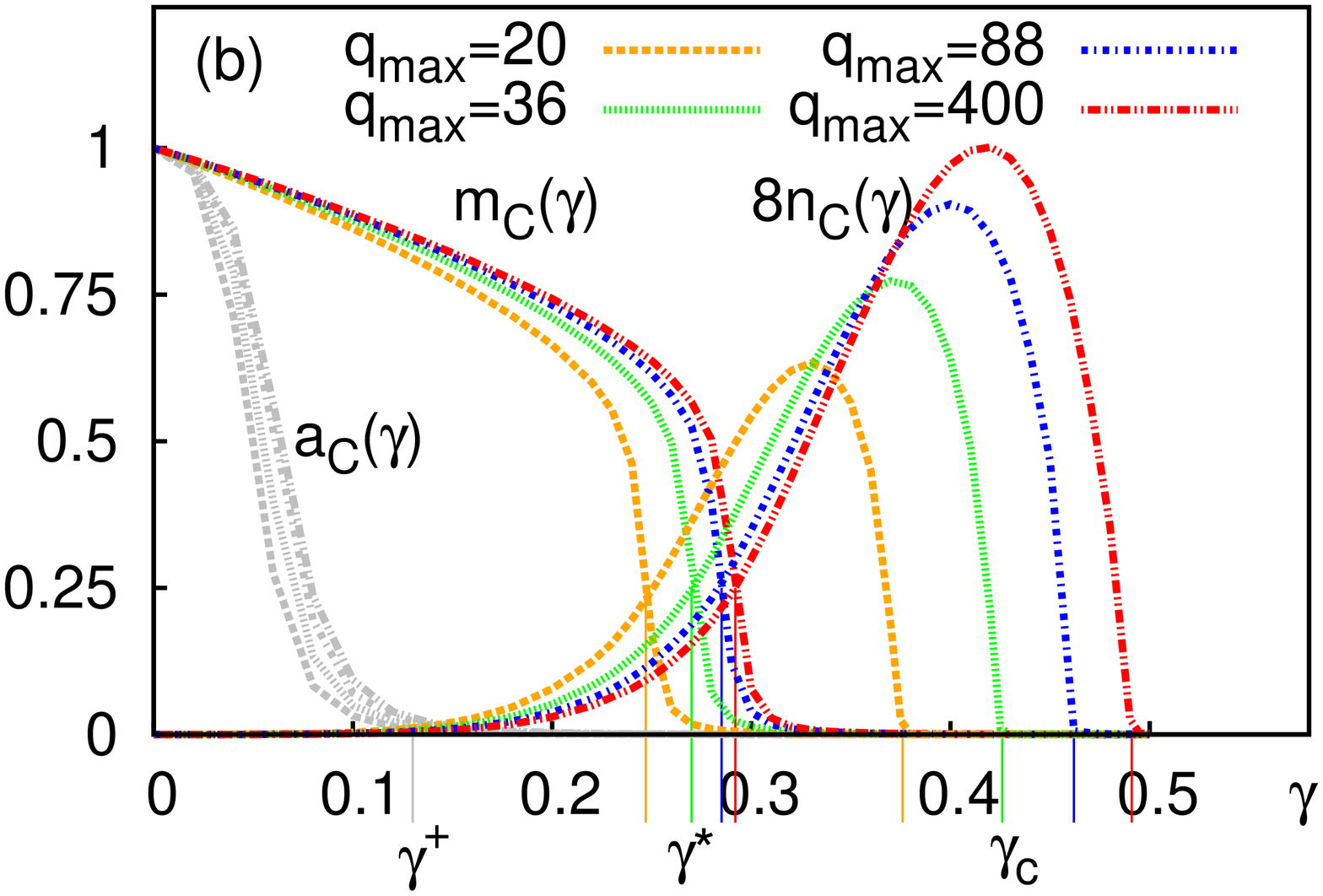}
\caption{\label{qfix_clustering}Reduced average cluster size $a_C(\gamma)$, maximum cluster
size $m_C(\gamma)$ and reduced number of clusters $n_C(\gamma)$ \vs\ curing rate
$\gamma$ in the stationary state. (a) CA $SIS$ model with fixed neighbourhood size $q$ indicated in the figure;
(b) CA $SIS$ model with random neighbourhood size at various $q\idx{max}$ indicated in the figure.
Characteristic values of $\gamma$ that separate various clustering regimes are denoted as
$\gamma^{+}$, $\gamma^*$ and $\gamma_c$ (critical curing rate).}
\end{center}
\end{figure}
\begin{figure}
\begin{center}
\includegraphics[clip,angle=0,width=12cm]{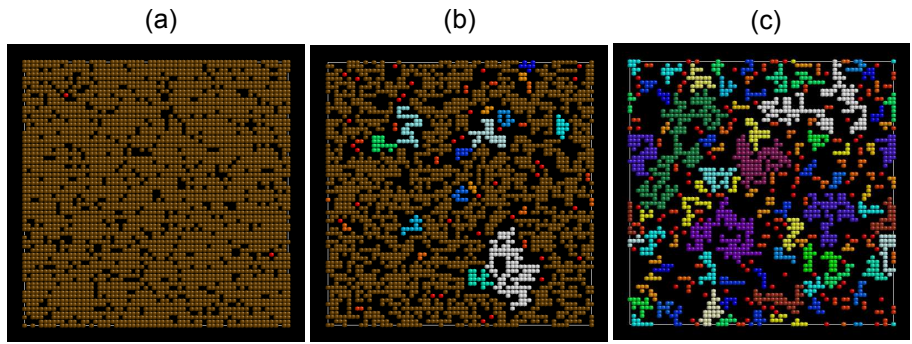}
\caption{\label{qfix_snaps}Snapshots with colour-coded clusters of infected
individuals for CA $SIS$ model with fixed neighbourhood size $q=4$.
Susceptible individuals are not shown. (a)-(c) illustrate respective clustering regimes
described in the text.}
\end{center}
\end{figure}

The behaviour of these averages is shown in Fig.~\ref{qfix_clustering} for both cases of
CA $SIS$ model with fixed $q$ [frame (a)] and CA $SIS$ model with random $q$ [frame (b)].
The set of values $q=4,8,20,96$ used for the model with fixed $q$ matches the respective set
of values $q\idx{max}=20,36,88,400$ for the model with random $q_k$ in terms of close
respective values of $\gamma_c$, as discussed in Sec.~\ref{IV}, Eq.~(\ref{qmax_q}).  
For both types of models one can distinguish three following regimes:
\begin{enumerate}[(a)]
\item single large cluster ($\gamma<\gamma^{+}$): $n_C(\gamma)\sim 0$,
$a_C(\gamma)>0$, $0.9<m_C(\gamma)<1$;
\item large + small clusters ($\gamma^{+}<\gamma<\gamma^{*}$):
$0<n_C(\gamma)<2$, $a_C(\gamma)\sim 0$, $0.25<m_C(\gamma)<0.9$;
\item small clusters only ($\gamma^{*}<\gamma<\gamma_{c}$):
$n_C(\gamma)$ peaks, $a_C(\gamma)\sim m_C(\gamma)\sim 0$.
\end{enumerate}
The value of $\gamma^+$ is defined approximately, whereas $\gamma^*$ is the inflection point for the
$m_C(\gamma)$. The regimes (a)-(c) are illustrated in the series of snapshots shown in
Fig.~\ref{qfix_snaps} that are obtained for the smaller system of $60 \times 60$ individuals.

The presence of these regimes indicates that besides the phase transition that occurs at
$\gamma\to\gamma_c$, where $I(\infty)$ vanishes, there is another transition which takes place earlier,
at $\gamma\to\gamma^*$, where the size of the largest cluster of infected individuals decreases down to zero.
At $\gamma^*<\gamma<\gamma_c$, only small disconnected clusters are to be found [regime (c) above].
The position of this transition, given by $\gamma^*$, is found to be weakly dependent on $q$ for the CA
$SIS$ model with fixed $q$ [Fig.~\ref{qfix_clustering} (a)]. However, despite matching both models \via\ their
respective values of $\gamma_c$, their respective values of $\gamma^*$ do not match.
This is especially evident for the cases of $q\idx{max}=20$ \vs\ $q=4$ and $q\idx{max}=36$ \vs\ $q=8$.
In the model with random implicit mobility ($q\idx{max}=20$ and $36$) the maximum cluster size
vanishes at lower curing rate $\gamma$ when compared to the model with fixed implicit mobility
($q=4$ and $8$). Therefore, the randomness in the neighbourhood size $q_k$ promotes stronger splitting
of infected individuals into separate clusters. The higher maximum values of $n_C(\gamma)$ occuring
within the interval $\gamma^*<\gamma<\gamma_c$ for the cases $q\idx{max}=20$ and $36$ as compared to
the respective cases $q=4$ and $8$ also supports this conclusion.

\begin{figure}
\begin{center}
\includegraphics[clip,angle=270,width=9cm]{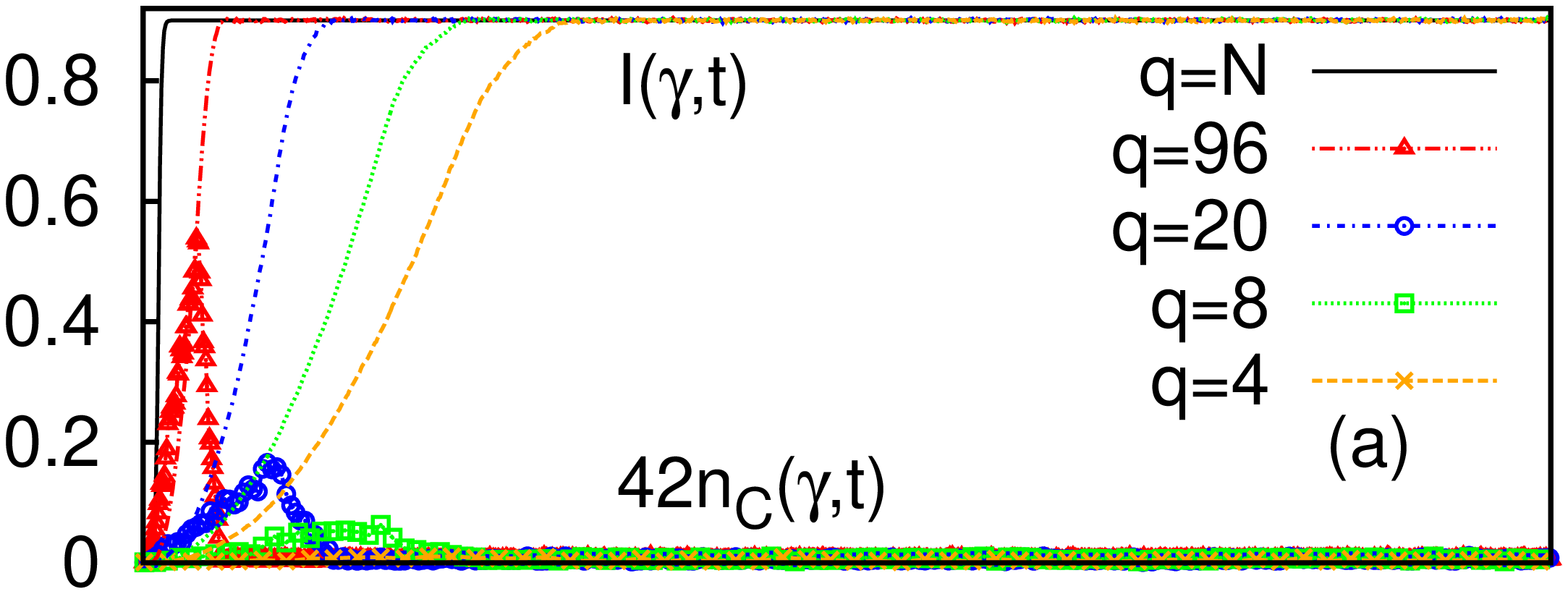}\vspace{-5mm}\\
\includegraphics[clip,angle=270,width=9cm]{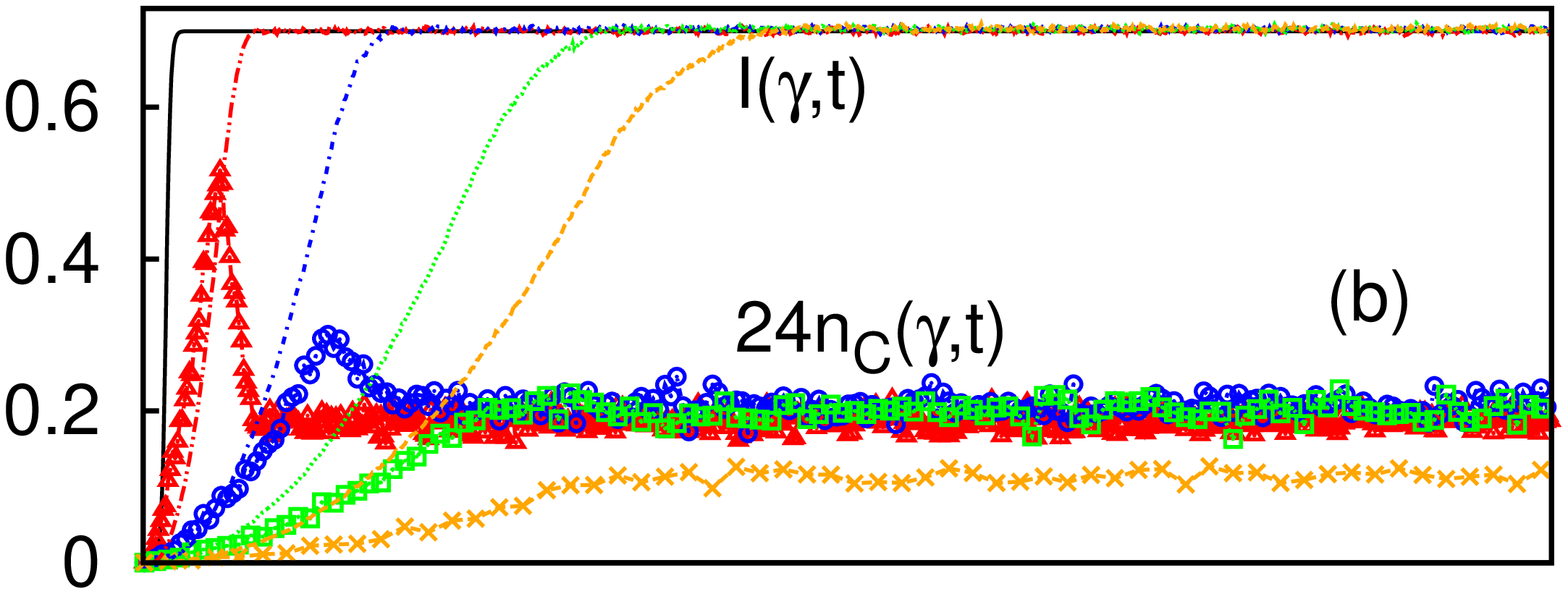}\vspace{-5mm}\\
\includegraphics[clip,angle=270,width=9cm]{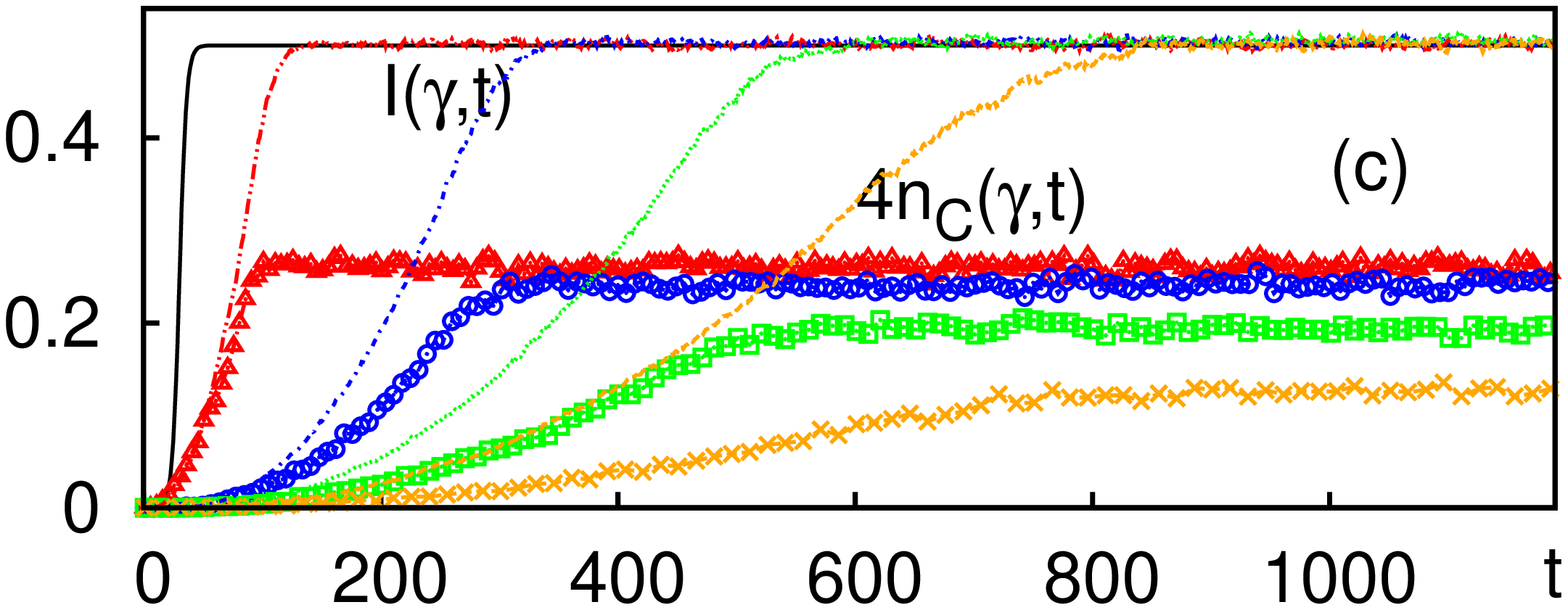}
\caption{\label{qfix_clust_evol_A}Time evolution of the reduced number of clusters $n_C(\gamma,t)$
and the fraction of infected individuals $I(\gamma,t)$ for the CA $SIS$ model with
fixed neighbourhood size $q$ indicated in the figure. (a) simulations at respective values of $\gamma$
that lead to $I(\infty)=0.9$ for each $q$; (b) the same for $I(\infty)=0.7$; (c) the same for $I(\infty)=0.5$.}
\end{center}
\end{figure}
\begin{figure}
\begin{center}
\includegraphics[clip,angle=270,width=9cm]{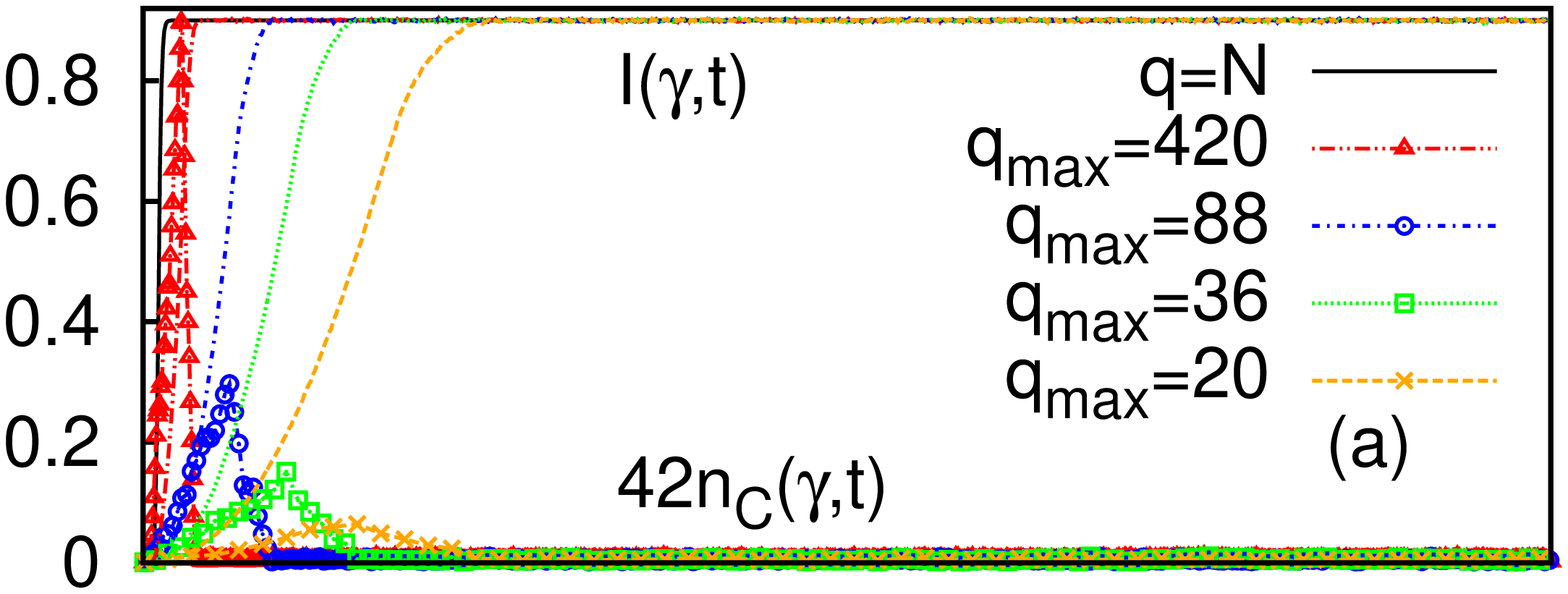}\vspace{-5mm}\\
\includegraphics[clip,angle=270,width=9cm]{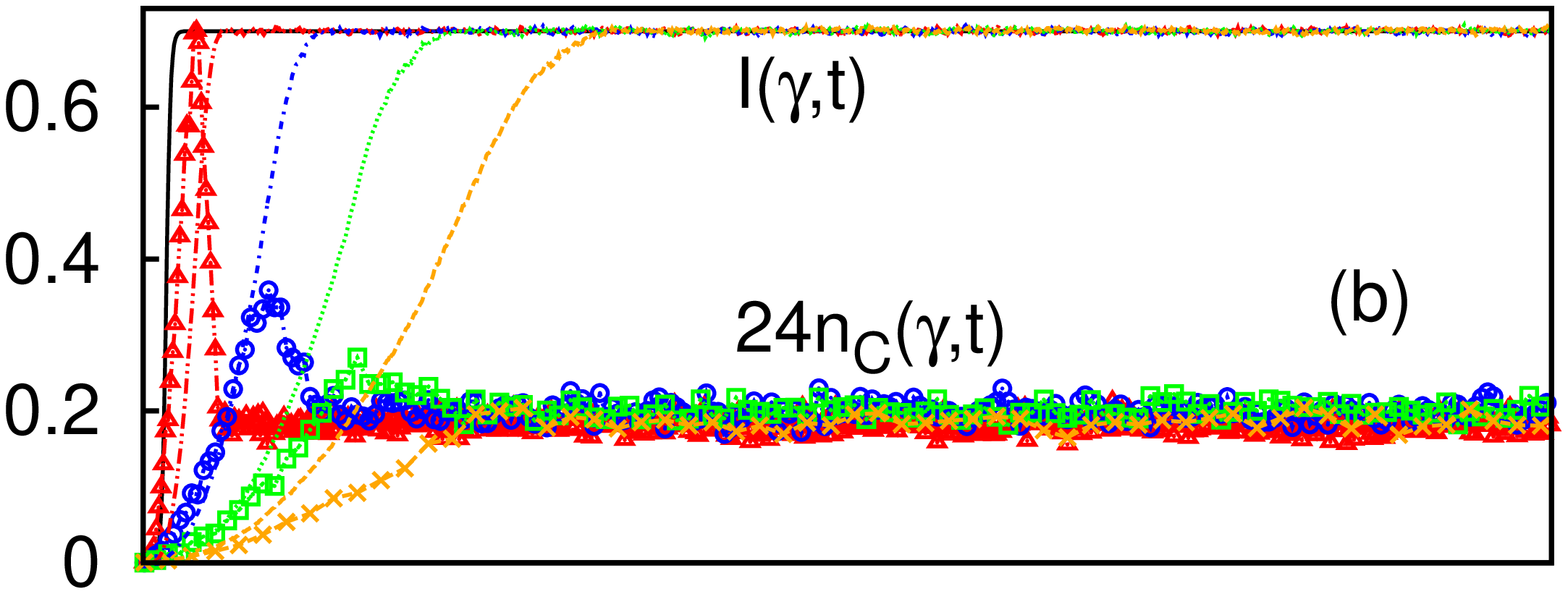}\vspace{-5mm}\\
\includegraphics[clip,angle=270,width=9cm]{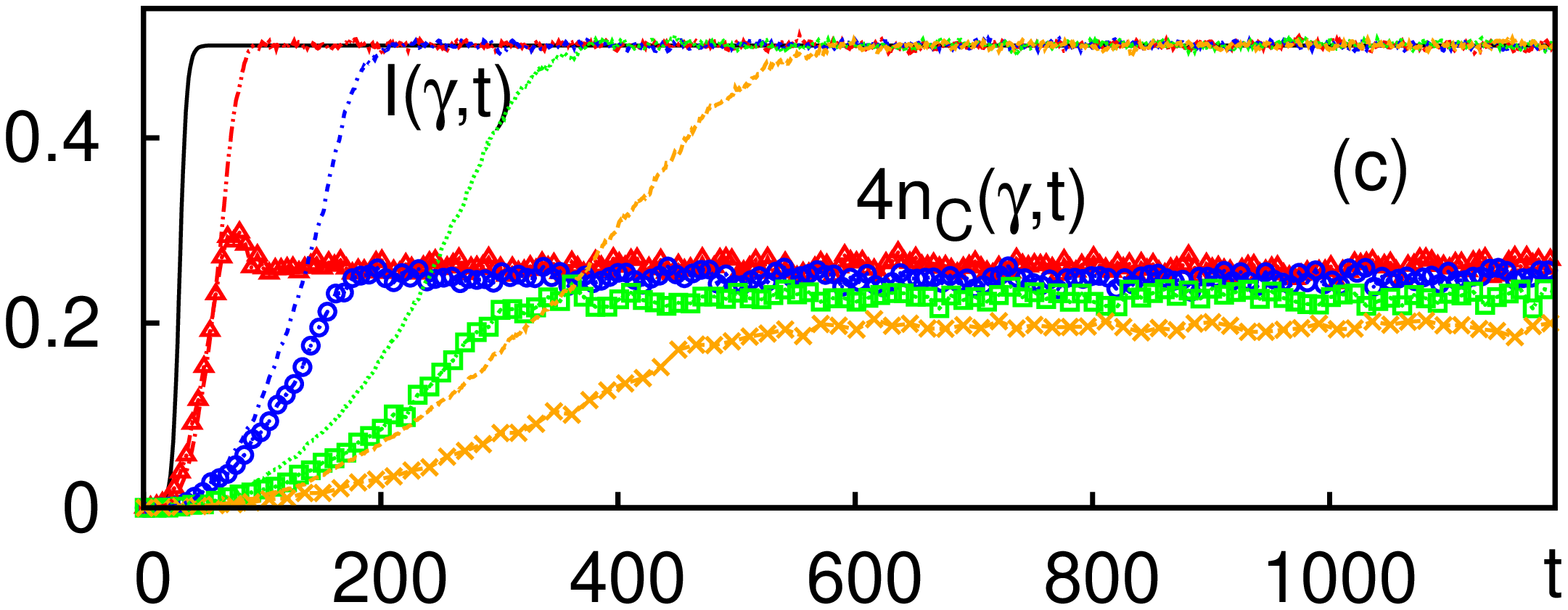}
\caption{\label{qfix_clust_evol_B}The same as in Fig.~\ref{qfix_clust_evol_A} except
for the CA $SIS$ model with random neighbourhood size $q_k \le q\idx{max}$ indicated in the figure.}
\end{center}
\end{figure}

We will consider now the evolution of relevant properties towards the stationary state in both models.
To this end the series of runs are performed each started from the initial configuration containing
a single infected individual. These series are split into three sets: (a) very low curing rate;
(b) low curing rate; and (c) moderate curing rate. Each set contains a number of runs performed for
both models at various values of $q$ and $q\idx{max}$, respectively. The curing rate $\gamma$ is chosen
individually for each run from the condition that the stationary state with $I(\infty)=0.9$, $0.7$,
and $0.5$ is reached when the run belongs to (a), (b), and (c) set, respectively.  
The evolution of the instant fraction of infected individuals $I(\gamma,t)$ and reduced
number of clusters $n_C(\gamma,t)$ are plotted \vs\ time $t$ for both models in Figs.~\ref{qfix_clust_evol_A}
and \ref{qfix_clust_evol_B}, respectively. The solution (\ref{SIS_red_solut}) for $I(\gamma,t)$ obtained
for the zero dimensional $SIS$ model is also shown and is marked \via\ $q=N$.

Within each set, (a), (b), or (c), for the model with fixed neighbourhood size $q$, the increase
of $q$ leads to the reduction of a timescale for the system to reach the stationary state,
monitored \via\ behaviour of $I(\gamma,t)$ (see, Fig.~\ref{qfix_clust_evol_A}). Such effect
is to be expected, as far as the increase of $q$ is interpreted
in terms of enhancement of the implicit mobility, which promotes faster spatial spread of the infection.
The fastest infecting dynamics is, obviously, achieved within the zero dimensional $SIS$ model
corresponding to the ideally mixed case
\cite{Boccara1992,Boccara1993,Boccara1994,German2011}.
Upon introducing random neighbourhood size $q_k$, the infecting dynamics also speeds-up when compared
to the results of respective matching runs performed at fixed $q$ (compare Figs.~\ref{qfix_clust_evol_A}
and \ref{qfix_clust_evol_B}). The speed-up ranges from about $1.5$ to $2$ times depending on the
particular set, (a), (b), or (c).

The evolution of spatial patterning also demonstrates marked dependence on the model type and the
parameters of the run. In particular, for the case of fixed $q$ and moderate curing rate 
[see, Fig.~\ref{qfix_clust_evol_A}, (c)], the reduced number of clusters $n_C(\gamma,t)$
approaches its stationary value monotonically from below. However, the behaviour is different at
very low and low curing rates [frames (a) and (b)]. In these cases, at larger neighbourhood sizes,
$q\geq 8$, there is a pronounced maximum in the shape of $n_C(\gamma,t)$ curve, approximately a
mid-way to the stationary state. With the increase of $q$ its height increases. The explanation for the
presence of this maximum is the follows. At the early stage the system is mostly uninfected
and for larger infection range the newly infected individuals will be formed at some
distance from existing infected ones thus increasing the number of separate clusters $n_C(\gamma,t)$.
However, at a mid-way to the stationary state about half of the system is already infected and newly
infected individuals begin to link the existing clusters, thus reducing the number $n_C(\gamma,t)$.
Similar behaviour is also seen within the model with random $q_k$ (shown in Fig.~\ref{qfix_clust_evol_B}),
where even higher maxima are observed when compared to matching cases of the model with fixed $q$.
Therefore, the introduction of the randomness in the neighbourhood size enhances the hill-like shape
in the evolution of the number of clusters $n_C(\gamma,t)$. At very low curing rate
[frame (a)] the peaks are observed for all $q\idx{max}$ being considered.

Therefore, the dynamics of spatial patterning for infected individuals is found to depend strongly
on the model type and the parameters of the run. When the initial state comprises almost
healthy system (save for one individual), then for the contact-process-like infection spread, $q=4$,
the grows of a number of infected clusters towards its stationary value is always monotonic.
With the increase of the neighbourhood size, $q>4$, this behaviour changes and the number of
clusters shows a pronounced maximum about a mid-way towards its stationary value. This type of behaviour
is getting more pronounced for the model with random neighbourhood size.

\section{\label{VI}Conclusions}

In the present work, we study the $SIS$ model on geometric graph obtained from 
$\mathbb{Z}^2$. To each vertex the individual is attached with two possible states:
susceptible or infected. Two options are considered for the neighbourhood size $q$, namely:
the fixed $q$; and that taken at random at each step and for each individual $q_k$ globally
bounded by $q\idx{max}$.
The neighbourhood size is interpreted as the level of individuals' implicit mobility.
The evolution of the system of $N$ individuals is run according to the asynchronous cellular
automata algorithm. 

The stationary states are studied first, where we concentrated on the phase transition
like behaviour for the ratio of infected individuals $I(\infty)$ which vanishes at the
critical curing rate $\gamma=\gamma_c$. For the model with fixed $q$, with the increase of $q$,
$\gamma_c$ increases towards the value $0.5$ characteristic for the zero dimensional
$SIS$ model, which corresponds to the case of $q=N$. The simulation data obtained at a range
of $q$ is used then to extend the solution for the zero dimensional model to cover the case
$q<N$. Simple power-law model form is found for $\gamma_c$ as the function
of $q$. Close match is found for the phase transition like behaviours of the models with
fixed $q$ and random $q_k$ bounded by $q\idx{max}$ when $q$ and $q\idx{max}$ obey certain
relation.

In the stationary state we found three distinct regimes of spatial patterning of infected
individuals depending on the curing rate $\gamma$: (a) single cluster; (b) large+small
clusters; and (c) small clusters only. The phase transition like change between (b) and (c)
occurs at $\gamma^*$, the inflection point for the size of the largest cluster. The value
of $\gamma^*$ is found to be weakly dependent on $q$ for the fixed neighbourhood size model
but markedly dependent on $q\idx{max}$ for the model with random neighbourhood size. It is
found that randomness in $q_k$ promotes splitting of the largest cluster at lower $\gamma$
as compared to the case of matched case with fixed $q$.

Both the increase of the neighbourhood size and its randomness reduce the time needed by the
system to reach the stationary state when simulation is started from the almost healthy state.
This is understood in terms of the relation between the neighbourhood size and the individuals'
implicit mobility. The type of evolution of spatial patterning also demonstrates marked dependence
on the model type and the parameters of the run. For the contact-process-like infection spread, $q=4$,
the grows of a number of infected clusters towards its stationary value is always monotonic.
With the increase of the neighbourhood size, $q>4$, this behaviour changes and the number of
clusters shows a pronounced maximum about a mid-way towards its stationary value. This type of
evolution, which involves the initial built of a huge number of separated clusters and their following
merge into larger ones, is getting more pronounced for the model with random neighbourhood size.

This study, performed for a relatively simple $SIS$ model can be extended towards more complex
models involving latency, immunity, other types of graphs and memory effects.

\section{Acknowledgements}

This work was supported by the International Research Staff Exchange Scheme grant
“Structure and Evolution of Complex Systems with Applications in Physics and Life Sciences”
STREVCOMS-612669 within the 7th European Community Framework Program.

\bibliography{biblio.bib}{}
\end{document}